# Graphene field-effect transistor array with integrated electrolytic gates scaled to 200 mm

N C S Vieira[1,3], J Borme[1], G Machado Jr.[1], F Cerqueira[2], P P Freitas[1], V Zucolotto[3], N M R Peres[2] and P Alpuim[1,2]

[1]INL - International Iberian Nanotechnology Laboratory, 4715-330, Braga, Portugal.
[2]CFUM - Center of Physics of the University of Minho, 4710-057, Braga, Portugal.
[3]IFSC - São Carlos Institute of Physics, University of São Paulo, 13560-970, São Carlos-SP, Brazil

E-mail: pedro.alpuim.us@inl.int

**Abstract**
Ten years have passed since the beginning of graphene research. In this period we have witnessed breakthroughs both in fundamental and applied research. However, the development of graphene devices for mass production has not yet reached the same level of progress. The architecture of graphene field-effect transistors (FET) has not significantly changed, and the integration of devices at the wafer scale has generally not been sought. Currently, whenever an electrolyte-gated FET (EGFET) is used, an external, cumbersome, out-of-plane gate electrode is required. Here, an alternative architecture for graphene EGFET is presented. In this architecture, source, drain, and gate are in the same plane, eliminating the need for an external gate electrode and the use of an additional reservoir to confine the electrolyte inside the transistor active zone. This planar structure with an integrated gate allows for wafer-scale fabrication of high-performance graphene EGFETs, with carrier mobility up to 1800 cm$^2$ V$^{-1}$ s$^{-1}$. As a proof-of principle, a chemical sensor was achieved. It is shown that the sensor can discriminate between saline solutions of different concentrations. The proposed architecture will facilitate the mass production of graphene sensors, materializing the potential of previous achievements in fundamental and applied graphene research.

## 1. Introduction

Since the first report of a graphene-based field-effect transistor (FET) over a decade ago [1], a number of FET-based sensors and biosensors using graphene, graphene oxide, and related graphene nanostructures have been developed for physical, chemical, and biological applications [2, 3]. Due to graphene's unique electronic properties, combined with its high chemical stability and structural uniformity, graphene FETs seem to be ideal platforms for the selective detection of molecules with relevance in many areas [4], ranging from disease diagnosis [5] to environmental monitoring [6] and security [7]. Graphene science and technology are currently undergoing a critical stage, in which graphene's outstanding properties, demonstrated in many research laboratories across the world, are put to test upon up-scaling to an industrial product, processed for human use. This step has not yet been achieved for most of the promised graphene applications [8]. In this context, no matter how big the potential advantages of graphene FETs for chemical and biological sensing are, their exploitation in real applications makes it obvious that issues related to a high level of device integration, device portability, high fabrication throughput, and reliability, must be addressed and overcome before mass production of graphene-based products becomes a reality.

From a biochemical point of view, it is a great advantage that a graphene FET displays equal or even improved performance when the solid-state gate dielectric is replaced by an electrolytic solution [9]. For this reason, the electrolyte-gated field-effect transistor configuration (EGFET) is the preferred choice for this purpose [10]. In addition, graphene EGFETs operate at lower gate voltage, because almost all the voltage applied to the gate electrode drops in the nanometer-sized electrical double layers (EDLs) that form at the gate/solution and solution/graphene interfaces [10]. This results in a much higher electrostatic capacitance per unit channel area than in conventional back-gated structures, where the gate contact and the graphene channel are separated by tens or hundreds of nanometers of a solid dielectric. A consequence of the higher capacitance of the

EDL in graphene EGFETs is that the quantum capacitance of graphene can no longer be ignored for device modeling, since both capacitances are of the same order of magnitude [11, 12].

When compared to other transistor architectures, EGFETs usually require the use of a large, cumbersome, gate electrode (generally a silver/silver chloride reference electrode or a metallic wire made of gold, platinum or silver) [9, 10], which represents a hindrance for miniaturization and integration. This feature may preclude the use of graphene EGFETs in applications like point-of-care testing (e.g. disposable biosensors), where a compact, integrated design is required. Another challenge is the potential for upscaling the technology, e.g., its suitability for fabrication at the wafer-scale, like conventional inorganic transistors.

In the present work, we report the fabrication, operation, and modeling of a fully-integrated graphene EGFET architecture, where the conventional wire gate electrode is replaced by an in-plane recessed metallic gate, which is replicated at the wafer scale by means of a standard UV-optical lithography clean-room process that is rendered compatible with graphene. The integrated gate geometry provides an efficient transistor gating and also confines the droplet inside the transistor active zone. The structure, including the pads and metallic lines connecting source, drain and gate electrodes, is then replicated 280 times in an array that covers the surface of a 200 mm oxidized silicon wafer. The single-layer graphene EGFETs resulting from this process consistently perform at the same level as that reported for devices based on exfoliated or CVD (chemical vapor deposition) graphene flakes transferred onto small-area substrates.

## 2. Results and discussion

### 2.1. Graphene EGFET architecture

Figure 1 shows optical images of a 200 mm wafer patterned with 280 transistors (figure 1a), a zoomed-in view of an individual device undergoing measurement (figure 1b), and a microscope image of the transistor's gold (Au) source, drain, and integrated gate contacts (figure 1c). The three pads visible in the foreground of figure 1b are, from left to right, for the source, drain, and gate contacts. The electrolyte droplet is clearly visible. The gray lines are intended to act as guides in the wafer dicing process. In figure 1c the inner lobe of the ring-shaped gate contact, with internal and external diameters of 200 μm and 1000 μm, respectively, is visible. The outer lobe (only partially visible) has internal and external diameters of 2000 μm and 3000 μm, respectively. The transistor architecture is that of a planar FET, with a recessed, ring-shaped gate placed in the same plane as the source and drain contacts and the graphene channel. This architecture differs from top-gate and bottom-gate architectures in that it does not contain a solid-state dielectric layer between the graphene channel and the metal gate. Here, in contrast, the graphene surface remains available to interact with the electrolyte solution. In the absence of the electrolyte droplet, the gate contact is electrically insulated from the transistor channel. A drop of solution provides the capacitance required to operate the graphene EGFET.

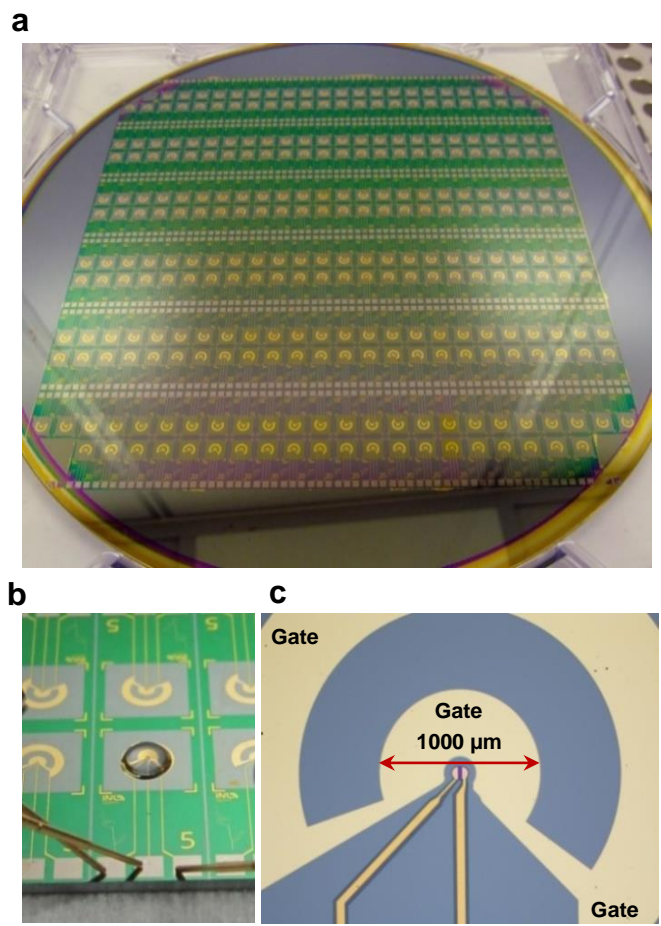

**Figure 1**. Images of graphene EGFETs: (a) 200 mm wafer patterned with 280 transistors. (b) A device with $W/L = 12$ (labelled 5) being measured. The pads for the source, drain and gate contacts, visible in the foreground, have sides of 1.5 mm. The gray, aluminum lines are guides for the wafer dicing process. (c) The transistor's Au source, drain and integrated gate contacts. The inner lobe of the ring-shaped gate contact, with internal and external diameters of 200 μm and 1000 μm, respectively. The outer lobe (only partially visible) with internal and external diameters of 2000 μm and 3000 μm, respectively.

In the current design, the two Au concentric circular zones connected at the edges that form the gate, are defined on a silicon dioxide ($SiO_2$) squared area (visible in figure 1b), at the center of which the graphene channel is patterned with overlap onto the source and drain Au contacts (figure 1c). This design provides a contrast in surface energy, $\gamma$, between the Au ($\gamma \sim 1.50$ J/m$^2$), and the $SiO_2$ ($\gamma \sim 0.287$ J/m$^2$) areas [13, 14], which helps confine the water droplet used as gate dielectric between the two gold concentric circular zones, perhaps with some overlap (depending on the volume of the droplet), to the external zone but not beyond. In figure 1c, the water droplet has volume of 5 μL.

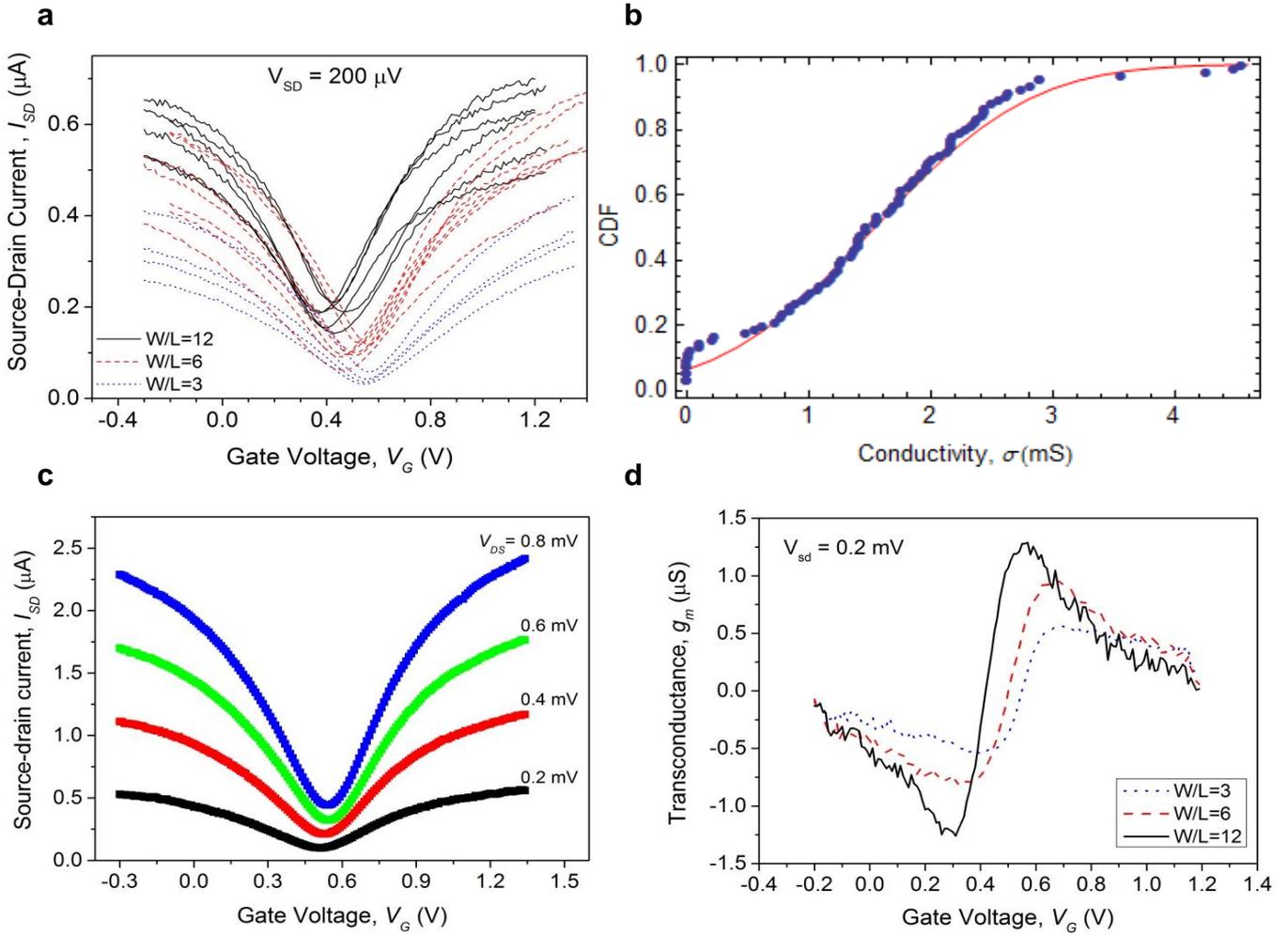

**Figure 2.** (a) Transistor transfer curves of 17 graphene EGFETs fabricated on a 200 mm wafer with $W/L = 3$ (blue dotted lines), 6 (red dashed lines) and 12 (black solid lines). PBS was used as the electrolyte-gate dielectric and $V_{SD} = 0.2$ mV. (b) Empirical CDF of sample conductivity data taken from a sample of 90 devices (blue solid circles), and CDF of the corresponding normal distribution (red line). (c) Transfer curves of an $L = 12.5$ μm transistor, for different values of $V_{SD}$ ($V_{SD} = 0.2, 0.4, 0.6$ and $0.8$ mV). (d) Transconductance, $g_m$, for three devices with $W/L = 3, 6$ and 12 obtained at $V_{SD} = 0.2$ mV.

Our proposed EGFET architecture has many advantages as compared to the usual design of EGFETs that use an external gate electrode and source-drain contacts which overlap the channel material. In the first place, it allows for the fabrication of integrated chips, with all the transistor contacts and the active layer placed side by side, and the respective pads placed along the chip edge. Such a design makes it very easy to insert the complete chip into a connector or to wire-bond it onto a PCB board. Moreover, from a fabrication point of view, this architecture has the advantage of gathering together all additive and subtractive lithographic steps related to the patterning of metallic and dielectric layers at the initial stages of the fabrication process, effectively dissociating these steps from the graphene process. This allows for a better process design, free from constrains that would emerge if performing sputter deposition, dry etching and lift-off in the presence of graphene. Furthermore, delaying all graphene-related steps as much as possible in the fabrication flow chart preserves the quality of the patterned graphene in the finished device.

## 2.2. Characterization of graphene EGFETs

In the fabrication process, the area of the graphene samples is limited to 100 mm × 150 mm by the size of the quartz tube and the paddle that holds the substrates inside of it. Therefore, we could not transfer graphene onto the entire pre-patterned 200 mm wafer in a single step. Hence, graphene was grown in two batches of multiple Cu foils and transferred until the desired coverage of the wafer was achieved. After patterning the graphene, a random sample of 90 devices was collected, and the source-drain resistance of the transistors was measured in air, without gating, for quality control. We considered the threshold value for rejection of devices to be 10 kΩ. This choice is arbitrary and was based on repeated measurements of graphene FETs, which showed that devices with high channel resistance (larger than 10 kΩ for the present geometry) did not survive more than a few repetitions of the electrical measurements. Eleven devices had resistance above the threshold and were rejected (12% of the sample size). The remaining 79 devices (88% of the sample) were tested in a 3-terminal configuration, all displaying very clearly graphene transistor behavior. Transfer curves in phosphate buffered saline (PBS) for 17 transistors with channel length 6.25 μm, 12.5 μm, and 25 μm, are displayed in figure 2a.

Channel width is 75 μm in all transistor configurations. $V_{SD}$ was fixed at 0.2 mV.

Figure 2b shows the empirical cumulative distribution function (CDF) of the conductivity data of 90 transistors, under no gate voltage, as solid blue circles

The conductivity is defined as:

$$\sigma = \frac{1}{R} \times \frac{L}{W} \qquad (1)$$

where $R$ is the resistance, and $L$ and $W$ are the channel length and width, respectively. Also shown for comparison (red line) is the CDF of a normal distribution with parameters $\mu$ and $\sigma_0^2$, numerically equal to the average of sample conductivity ($\bar{\sigma}$ =1.537 mS) and to the sample conductivity variance ($s^2$ = 1.028 mS$^2$), respectively. For conductivities $\gtrsim$ 0.6 mS, the empirical CDF closely follows the normal distribution. For values of conductivity close to zero, the empirical distribution is very different from the normal distribution. This is because, in that particular range, experimental data correspond to the accumulation points for all fully and partially-broken transistor channels (conductivity ~ 0 S).

In figure 2a, the graphene EGFET transfer curves, i.e., the drain current as a function of gate voltage, were taken under a constant source-drain voltage ($V_{SD}$ = 0.2 mV), using PBS solution as the electrolyte-gate dielectric. The curves display the typical features of graphene transistors [1], i.e., the conductivity is modulated by the gate voltage in a symmetric way around a point of minimum conductivity, which corresponds to the positioning of the Fermi level at, or close to, the Dirac point. The two branches of the curve, to the left and right of the conductivity minimum, correspond to transport by holes and electrons, respectively. From the position of the Dirac point, which was always found to be shifted towards positive values of $V_G$, it was concluded that the graphene was unintentionally p-doped. This is a common feature observed in CVD graphene devices, which can be attributed to polymer residues (photoresist and PMMA) [15], doping due to water/oxygen adsorbed at the graphene surface [16], or, in areas where they are present, to the metal contacts underneath [17].

Two trends are evident in the series of transfer curves in figure 2a. One is a shift upwards, towards higher $I_{SD}$, as $W/L$ increases. This is explained by the smaller channel resistance as the channel length becomes shorter, at similar doping levels. The second observation is the shift of the minimum conductivity point towards lower values of $V_G$ as $W/L$ increases. This is tentatively explained by the asymmetry between the electrode areas in the gate's electrolytic capacitor system formed between the Au gate contact and the graphene channel. (See discussion ahead, when we introduce the equivalent capacitance for the circuit, and also the Supplementary Information).

Figure 2c shows the transfer curves in PBS of a graphene EGFET ($W/L$ = 6), for different values of constant source-drain voltage: $V_{SD}$ = 0.2, 0.4, 0.6 and 0.8 mV. It is clear from figure 2c that there is a broad operating range, allowing the transistor to operate at various power settings. Power is defined as $I_{SD} \times V_{SD}$, where the $V_{SD}$ is varied. The transistor could be operated at low power and thereby reduced transconductance (e.g. at $V_{SD}$ = 0.2 mV, $g_m$ = 1.0 and 0.81 μS for electrons and holes, respectively), or at high power and the corresponding enhanced transconductance (e.g. at $V_{SD}$ = 0.8 mV, $g_m$ = 4.5 and 3.8 μS for electrons and holes, respectively). Graphene EGFETs with high $g_m$ values (tens of μS) can be found in the literature (see, for example, Ohno et. al. [18] and Dankerl et. al [19]); however, those values are obtained at a very high $V_{SD}$ (~100 mV), approximately 100-500 times higher than the range used in the characterization of the EGFET proposed here. In this paper we focus on the low-power operating range of the graphene EGFETs since it insures that no voltage-induced chemical or biochemical reactions occur at or close to the active area of the device. It also extends the lifetime of the devices. In this operating range, the power consumption varies from 0.1 to 1 nW. However, this is done at the expense of having a low $g_m$. It is clear that operation at much higher $V_{SD}$, e.g. in the range of hundreds of mV, is possible, and it would result in correspondingly higher values of $g_m$. Figure 2d shows the $g_m$ of three devices with $W/L$ = 3, 6 and 12, obtained at low $V_{SD}$ (0.2 mV). Transconductance, defined as $g_m = dI_{SD}/dV_G$ (for $V_{SD}$ = constant), was calculated from the numerical derivative of the transfer curve, followed by a moving average filtering step.

The EGFET gate-drain leakage current is very low, in the range of 1-10 nA (see figure S2). For comparison, selected devices were gated using an Au wire, giving transfer curves very similar to those obtained with the integrated gate (figure S2). The leakage current using the wire gate was still very low, but higher than in case of the integrated gate.

Raman analysis of the device's channel area after all patterning steps were accomplished showed that the channel consisted of a single layer of graphene (SLG). Figure 3a and b show an optical microscope image and a typical Raman map, respectively, of one graphene

EGFET ($W/L = 6$). The Raman map in figure 3b has 9900 pixels, each containing a full Raman spectrum, acquired by a large area (110 × 90 μm$^2$) scan, with a resolution of 110 points per line and 90 lines, followed by a 3-cluster analysis. For details of the acquisition and interpretation of the Raman map, see the Supplementary Information.

It is clear from figure 3c that average spectra #1 and #2 are typical of SLG [16, 20], and that the channel region is essentially represented (having uniform color) by the average spectrum #1. The source and drain regions are fully covered with graphene that appears in some pixels to bear more resemblance to average spectrum #1, in others to average spectrum #2 and, in others yet, to a linear combination of both. The average spectra #1 and #2 differ mainly in the luminescent background that is observed as a drift in the baseline, and may be attributed to reflection of the laser light on the Au contacts. Average spectrum #3 has no graphene features and is associated with the Al$_2$O$_3$ covered areas.

### 2.3. Extracting graphene EGFET performance parameters

Conventional FET operation is based on the charging and discharging of a geometric capacitor (capacitance $C_g$) that is formed between the gate and the channel of the device, upon applying a gate voltage, $V_G$. In graphene transistors, another capacitance in series with the geometric one, called the quantum capacitance, $C_q$, must in certain cases be considered [11, 12], due to the vanishingly small density of states (DOS) of both the conduction and valence bands in the vicinity of the Dirac point:

$$V_G - V_{Dirac} = ne\left(\frac{1}{C_q} + \frac{1}{C_g}\right) \quad (2)$$

where $n$ is the carrier concentration in the transistor channel and $e$ is the elementary charge. The term $V_{Dirac}$ in equation (2) is the value of gate voltage for which the minimum $I_{SD}$ in the transfer curve of the device is observed. It accounts for possible unintentional doping of the graphene channel. In normal semiconductors, $C_q$ is very high when compared to the geometric capacitance, and therefore is negligible in equation (2). This is because the DOS at the semiconductor band edges is much higher than it is in graphene, where it is close to the Dirac point. In back-gated transistors the geometric capacitance is that of a parallel plate capacitor and is easily calculated by the following equation [19]:

$$C_g = \frac{\varepsilon \varepsilon_0}{d_{ox}} \quad (3)$$

where $\varepsilon$ is the dielectric constant of the gate dielectric ($\varepsilon = 3.9$ for SiO$_2$), $\varepsilon_0$ is the permittivity of free space, and $d_{ox}$ is the thickness of the gate dielectric. For SiO$_2$ with a typical thickness of 100 nm this gives $C_g \sim 35$ nF/cm$^2$. On the other hand, the quantum capacitance of graphene is [11]:

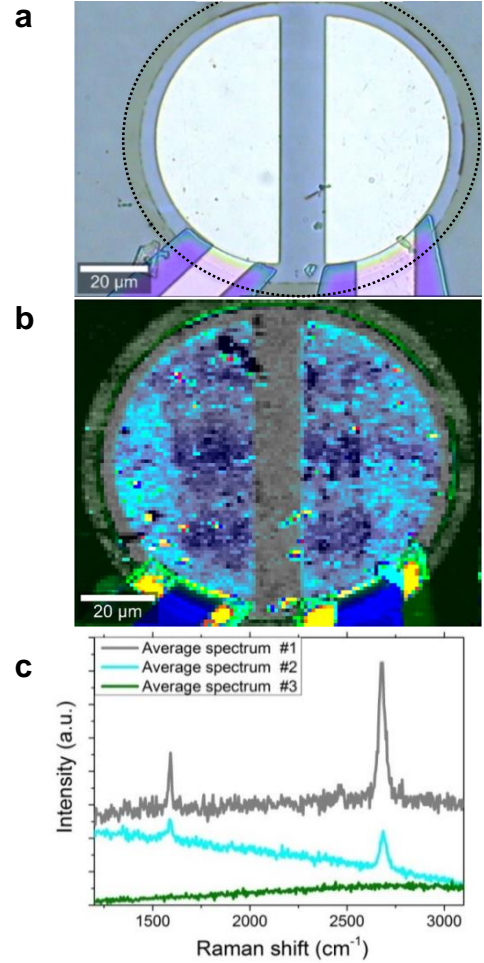

**Figure 3.** (a) Optical image of a typical graphene EGFET active area ($W/L = 6$), showing the brown-edged (resulting from overlap with Al$_2$O$_3$) quasi-circular graphene island (inside the dotted line added as a guide to the eye), the rectangular channel (blue) defined by the semicircular source and drain gold electrodes (white), and the surrounding SiO$_2$ background (gray). (b) Raman map of the same area as in (a) acquired with a large area (110 × 90 μm$^2$) scan. The colors result from applying a 3-cluster basis analysis to the image spectral data. (c) Average Raman spectra of each of the clusters used as basis to construct the image in (b). Calculations were made with Witec software Project FOUR+.

$$C_q = \frac{2e^2}{\pi}\frac{\sqrt{\pi n}}{\hbar v_F} \quad (4)$$

where $\hbar$ is Planck's constant and $v_F = 1.1 \times 10^8$ cm/s is the Fermi velocity. At a moderate doping level of $n \sim 5 \times 10^{12}$ cm$^{-2}$, equation (3) gives $C_q \sim 3$ μF/cm$^2$, which shows that for a back-gated graphene FET, the term $C_q$ can again be neglected in equation (2). Carrier concentration in this instance is given by:

$$n = \frac{V_G - V_{Dirac}}{e}C_g \quad (5)$$

A different situation arises in case of a graphene EGFET, where the geometric capacitance is that of the EDLs that form at the interfaces between graphene and electrolyte and between gate electrode and electrolyte, $C_{EDL}$. The thickness of this layer, $d_{EDL}$, is the Debye length, typically one to several nanometers [21], which is much smaller than the thickness of the dielectric in a bottom-gate graphene FET. This makes $C_q$ and $C_g = C_{EDL}$ of the same order of magnitude, and therefore both terms have to be considered in equation (2), e.g. for extracting $n$ as a function of $V_G$, a quantity that is critical for assessing the transistor performance.

However, direct measurement of $d_{EDL}$ is not readily accessible. Even a rough estimation of $d_{EDL}$ using Debye-Hückel theory [22] can be incorrect since the dielectric constant of water in very close proximity to a hydrophobic surface is different (smaller) than in the bulk [23], and so an accurate number to enter in the Debye-Hückel equation is missing. To continue studying the graphene EGFET, we therefore used a different approach that consists of fitting the transistor conductivity data using a theoretical model that does not rely on the use of equation (2). The model describes the dc conductivity of SLG, $\sigma$, as a function of the position of the Fermi level, based on carrier resonant scattering due to strong short-range potentials originating from impurities adsorbed at the graphene surface [24]. We use here a version of this theory adapted for the case when the carrier concentration, $n$, in graphene (i.e. the Fermi level

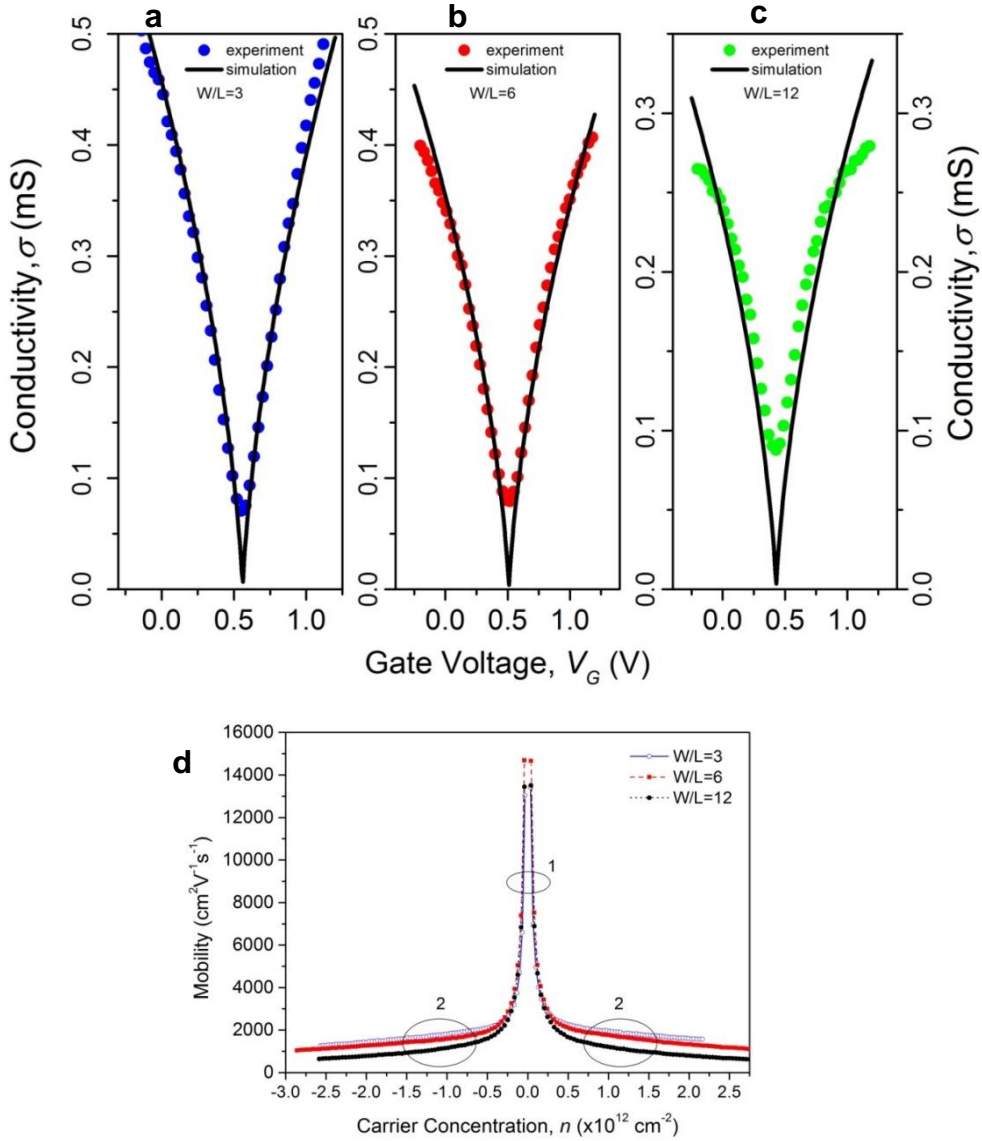

**Figure 4.** Graphene conductivity as a function of gate voltage for transistors with dimensions: (a) $W/L = 3$; (b) $W/L = 6$; (c) $W/L = 12$. Solid symbols are experimental data. Lines are the result of simulations to fit the data using the model described by equation (6). (d) Carrier mobility as a function of carrier concentration for devices with $W/L = 3$ (solid line), 6 (dashed line), and 12 (dotted line), respectively. Electron ($n < 0$ cm$^{-2}$) and hole ($n > 0$ cm$^{-2}$) branches are shown.

position) is set by a gate voltage [25]. In this particular form, the model reads:

$$\sigma = g_0 \frac{3\sqrt{3}}{4\pi} \frac{a_0^2 \alpha}{n_i} |V_G - V_{Dirac}| \ln^2(\sqrt{\alpha\pi}|V_G - V_{Dirac}| \cdot r) \quad (6)$$

where $g_0 = 2e^2/h = 0.078$ mS is twice the quantum conductance, $a_0 = 1.42$ Å is the C-C bond length in graphene, $\alpha = n/V_G$ represents an idealized capacitance when multiplied by the elementary charge, $n_i$ is the defect density, and $r \sim a_0$ is the range of the short-range potential created by the scattering centers. The fitting of the experimental data was done by finding numerical values for $r$, $\alpha$, and $n_i$ to which equation (6) gives a best fit of the conductivity data plotted as a function of the gate voltage. Some constrains were imposed on the range of values of the parameters, in order to ensure the physical significance of the solutions: $a_0 \leq r \leq 2a_0$, $0.5 \times 10^{12} \leq \alpha \leq 4.1 \times 10^{12} \text{FC}^{-1}\text{cm}^{-2}$, $5 \times 10^{11} \leq n_i \leq 2.5 \times 10^{12} \text{cm}^{-2}$. These numbers are obtained from the theoretical analysis of the conductivity curves of exfoliated graphene FETs on $SiO_2$ [25].

The quality of the observed fits (figure 4) shows that the chosen bounds are physically meaningful. We note that both the geometric and the quantum capacitances are included in the parameter $\alpha$, and cannot be disentangled. The optimum fitting parameters are listed in Table 1. The experimental conductivity data were extracted from the transfer curve of the devices according to equation (1).

Figure 4 shows the experimental data for the graphene conductivity as a function of $V_G$ (solid symbols), for graphene EGFETs with different W/L ratio: (a) W/L = 3; (b) W/L = 6; (c) W/L = 12. Figure 4 also shows, as continuous lines, the fitting of the data using equation (6). For the graphene EGFETs with W/L = 3 and 6, the fits are in very good agreement with the experimental data. For W/L = 12, the fit is not as good. Observation of the latter device with an optical microscope showed a graphene channel with many more wrinkles and vestiges of the clean room processing than were observed on the other two devices (figures 4a and b).

In the vicinity of the minimum conductivity point, the experimental data are not well fitted by the model. This is because the transport in graphene contacting the surface of a substrate (in our case, $SiO_2$) is governed by long-range potential fluctuations that give rise to the formation of electron and hole puddles [26], which are responsible for the finite conductivity of graphene at zero average carrier density. This type of interaction with the substrate is not accounted for in the model described by equation (4), which only communicates the transport physics in graphene at finite electronic densities [24]. The densities of scattering centers resulting from the simulations were $n_i = 1.8 \times 10^{12}$, $1.4 \times 10^{12}$, and $0.97 \times 10^{12}$ cm$^{-2}$ for the devices with W/L = 12, 6 and 3, respectively. The same value of $n_i$ was used to fit both the electron and hole branches of each curve.

Once carrier density as a function of $V_G$ is known, carrier mobility, $\mu$, can be calculated. This quantity provides a measure of the electronic quality of the graphene EGFETs. Figure 4d shows the Drude mobility as a function of carrier concentration, calculated from the conductivity data using the expression $\mu = \sigma(E_F)/en$, noting that for electrons $n < 0$ cm$^{-2}$ and for holes $n > 0$ cm$^{-2}$. Two distinct regions are clearly discernible in figure 4d. In region 1, corresponding to the neighborhood of the minimum conductivity point of graphene, both the electron and hole branches of the curves asymptotically increase as the average carrier concentration approaches zero, a value that experimentally is not accessible due to electron or hole puddles that form at the graphene/substrate interface (see discussion above). Region 2 corresponds to $|n| > \sim 0.5 \times 10^{12}$ cm$^{-2}$, is easily accessible experimentally, and is the region where most of the data points of the transistors' transfer curves fall. In particular, the linear regions in the curves of graphene conductivity as a function of $V_G$ (figure 4a-c) fall within this region, with a carrier concentration in the range of $2.5 \times 10^{11} \lesssim n \lesssim 1 \times 10^{12}$ cm$^{-2}$. From region 2 in figure 4d, it is seen that the transistor with W/L = 12 has lower mobility than the devices with W/L = 6 and 3. This is consistent with the large amount of residues observed by the microscope on the device surface, which in turn might be related to the poorer fitting in figure 4c to the model described by equation (6), as compared with the transistors of other dimensions. Moreover, the shorter the channel, the higher the influence of the contact resistance in the transistor curves [20], which would explain the reduced mobility at high carrier concentration in shorter channel devices. The device with W/L = 3 has the highest mobility, both for electrons and holes.

The most interesting parts of the transfer curves of the graphene EGFETs for sensing applications are possibly the almost-linear regions that lie to the right and left of the minimum conductivity point, in the electron and hole branches of the curves, respectively. There, the slope of each curve, $g_m$, is at its maximum, allowing for a maximum in device sensitivity. The shape of the curve ensures linearity. These regions correspond to moderate carrier densities with slowly varying carrier mobility as a function of carrier concentration, corresponding to the plateaus for $|n| \gtrsim 0.5 \times 10^{12}$ cm$^{-2}$ in figure 4d.

**Table 1.** Graphene EGFET performance parameters after fitting equation (4) to the experimental data.

| W/L | $V_{Dirac}$ (V) | $n_i$ (× $10^{12}$ cm$^{-2}$) | $\alpha$ (× $10^{12}$ F C$^{-1}$ cm$^{-2}$) | $\mu_h$ (cm$^2$ V$^{-1}$ s$^{-1}$) | $\mu_e$ (cm$^2$ V$^{-1}$ s$^{-1}$) |
|---|---|---|---|---|---|
| 12 | 0.43 | 1.77 | 4.1 | 768 | 794 |
| 6 | 0.51 | 1.37 | 4.0 | 1042 | 1224 |
| 3 | 0.56 | 0.974 | 3.4 | 1833 | 1843 |

We use the field-effect mobility equation for a FET [18]:

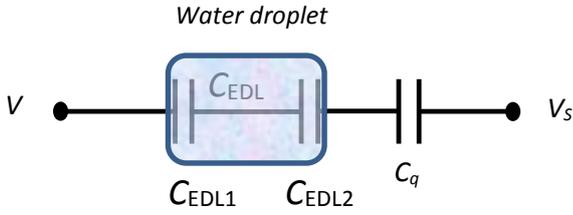

**Figure 5.** Diagram of the series capacitors associated with the gating circuit of the graphene EGFET. $C_{EDL}$ is the electrolytic gate capacitance provided by the water droplet, consisting of the series association of $C_{EDL1}$ and $C_{EDL2}$, which are the electrical double layer capacitances formed at Au/solution and solution/graphene interfaces, respectively. $C_q$ is is the quantum capacitance of graphene.

$$\sigma = \frac{L g_m}{W C_G V_{SD}} \quad (7)$$

to extract $\mu_{FE}$ from the graphene EGFET transfer curve where $C_G$ is gate capacitance. We start by evaluating $g_m$ at the inflexion points of the transfer curve, which appear as the extrema in figure 2d. For each graphene EGFET, a value of the parameter $\alpha$ in equation (6) resulting from the simulations (see Table 1) is used to calculate $C_G = \alpha$ $e$. Substituting $L$, $W$, $g_m$, $C_G$ and $V_{DS}$ in equation (7), we obtain values for $\mu_{FE}$ for both types of carriers ($\mu_e$ and $\mu_h$ for electrons and holes, respectively), in the range of ca. 750 – 1850 cm$^2$ V$^{-1}$ s$^{-1}$.

A remarkable feature of our graphene EGFET transfer curves is the high degree of symmetry of the electron and hole branches (e.g. a single value of $n_i$ fits both branches), yielding similar values of $\mu_e$ and $\mu_h$, as summarized in Table 1. This has been attributed to the Coulomb screening effect of the ions in the liquid electrolyte, neutralizing the charged impurities on the graphene surface originating from the SiO$_2$ substrate, which causes the scattering by impurities to be independent of carrier type [12, 27, 28].

One trend that is visible in the data (figure 2a) is that, on average, shorter channel devices have a $V_{Dirac}$ that is shifted to lower voltages. This might be a consequence of the asymmetry between electrode areas in the liquid gate capacitor, since the Au gate electrode has a fixed area for all devices, whereas the channel area depends on $L$ ($W$ is fixed). Since the two EDLs that form at the electrolyte/solid interfaces establish a capacitive voltage divider (figure 5), when the second capacitor decreases its area (equal to the channel area), the voltage drop across its terminals increases for the same $V_G$ across the series combination of both, thereby slightly increasing the charge concentration per unit channel area.

We estimate the magnitude of this effect, assuming reasonable values for all quantities involved (see the Supplementary Information), to be on the order of a of 10 ppm increase in channel charge concentration when going from devices with $W/L = 3$ to devices with $W/L = 12$. This change in carrier concentration is minute and is not enough to explain the shift in $V_G$ observed. Therefore, there must be other effects, possibly associated with the contacts, which also contribute to this shift. However, charge transfer from the Au contacts to graphene would lead to asymmetric transfer curves for electrons and holes [29], which is clearly not the case in our data. This effect requires further investigation.

### 2.4. The effect of the ionic strength of the electrolyte in device gating

The study of the response of the graphene EGFET to changes in the ionic strength of the gate electrolyte is relevant for biosensing applications, as biomolecules may come in a variety of aqueous solvents. To that end, a device having W/L = 6 was successively gated using aqueous solutions of NaCl with increasing concentrations: [NaCl] = 1.5, 15 and 150 mM, respectively. Figure 6 shows the transfer curves obtained for that device. It is evident that the transfer curves shift to lower $V_G$ as the electrolyte's ionic strength increases. This shift is -0.08 V per decade of ionic strength concentration. Electron and hole branches of the curves are symmetric around $V_{Dirac}$, giving similar $\mu_e$ (~ 1400 cm$^2$ V$^{-1}$ s$^{-1}$) and $\mu_h$ (~ 1300 cm$^2$ V$^{-1}$ s$^{-1}$). The source-drain current levels at same carrier concentration are similar in all cases

To better understand the family of transfer curves in figure 6, we consider that, for a constant level of $I_{SD}$, the shift in gate voltage is entirely due to a change in the capacitance of the device (figure 5) due to the different ionic strengths of the electrolyte. Since the transfer curves are similar (they are only shifted in the horizontal axis), a particular value of $I_{SD}$ taken in one curve corresponds, in the next one, to a shift in the horizontal axis $\Delta_1 = \Delta V_{Dirac} \sim 0.08$ V (see figure 6). Moreover, for constant $I_{SD}$, i.e., for constant carrier concentration, $C_q$ is constant. We can relate $n$ to the chemical potential of the graphene surface relative to the bulk of the solution, using Grahame's theory of the electrical double layer for a monovalent salt [30, 31]:

$$\rho = \sqrt{8\varepsilon\varepsilon_0 k_B T I} \sinh(\frac{e\psi_0}{2k_B T}) \qquad (8)$$

where $\rho = ne$ is the surface charge density, $I$ is the ionic strength of the electrolyte, $\varepsilon$ is the dielectric constant of water, $\varepsilon_0$ is the permittivity of free space, $k_B$ is the Boltzmann constant, $T$ is the absolute temperature, and $\psi_0$ is the surface potential. Since the dielectric constant of water at the hydrophobic graphene interface is not known, we leave it as a free parameter in equation (8). Substituting all known values of the parameters in (8) and solving for $\psi_0$, we obtain $\psi_0$ as a function of $\varepsilon$ for the ionic strengths studied. By fixing a physically realistic domain for $\varepsilon$ values, $5 \leq \varepsilon \leq 80$, one gets the corresponding intervals of absolute values of $\psi_0$: 1.60 V $\geq |\psi_0| \geq$ 1.53 V; 1.54 V $\geq |\psi_0| \geq$ 1.47 V; 1.48 V $\geq |\psi_0| \geq$ 1.41 V, when $I$ = 1.5, 15 and 150 mM, respectively. We can see that the values of $\psi_0$ shift towards lower voltages as $I$ increases. For a 10-fold increase in $I$, at any fixed value of $\varepsilon$, we find the shift $\Delta|\psi_0|$ = -0.06 V. This is the same trend and the same order of magnitude of the shifts observed in $V_G$ ($\Delta V_G \sim$ -0.08 V). Therefore it is possible to explain figure 6 based on the changes that occur in the EDL as a function of electrolyte's ionic strength. This is relevant to applications of the graphene EGFET as a chemical or biosensor.

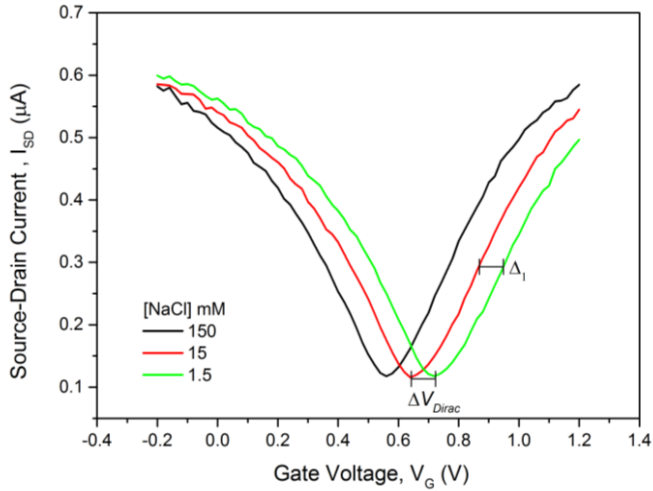

**Figure 6.** Transistor transfer curves for one graphene EGFET ($W/L = 6$) obtained at different ionic strengths of the gate electrolyte ([NaCl] = 1.5, 15 and 150 mM). The shift between transfer curves for measurements at 1.5 and 15 mM, $\Delta_1$, is highlighted: $\Delta_1 = \Delta V_{Dirac} \sim$ -0.08 V. The shift is similar between the curves measured at 15 and 150 mM.

An important step for future applications of our graphene EGFETs in sensing platforms is their portability and ease of use. For this reason, we designed and fabricated a printed circuit board (PCB) to support the devices, simultaneously providing easy, rugged, and precise electrical connections to the sourcing and measuring equipment (see figure S3 and supplementary information).

## 3. Conclusion

We demonstrated that graphene electrolyte-gated field-effect transistors can be integrated at the chip level by using a new transistor architecture, showing that the process can be up scaled to wafer-size microfabrication using standard clean-room processes. Our transistor architecture is based on a co-planar source, drain, and gate geometry, implying a recessed gate position relative to the active transistor channel region. Once the liquid gate electrolyte is added to the device the gate circuit is complete. The gold recessed gate is designed in such a way that it effectively confines the aqueous electrolyte in the transistor active area, ensuring that it will not spread over the chip. This graphene EGFET architecture lends itself to the use of microfluidics to release the electrolyte (possibly carrying an analyte) over the transistor channel. Transistors with channel length 25 μm showed an average field-effect electron mobility of 1500 cm$^2$ V$^{-1}$ s$^{-1}$ and average hole mobility of 1450 cm$^2$ V$^{-1}$ s$^{-1}$. Raman analysis of the transistor channel revealed that it consisted essentially of a single-layer graphene.

A model based on resonant scattering due to short-range potentials originated in impurities adsorbed at the graphene surface accurately fits the conductance data of the graphene EGFET in a broad range of gate voltage, especially at the approximately linear regions of the conductance curve that are most relevant for use of the graphene EGFET as a sensor. The transfer curve of the devices shifts $\sim$ -0.08 V for every 10-fold increase in ionic strength of the gate electrolyte, in the range 1.5 to 150 mM of NaCl. Based on the knowledge of carrier concentration extracted from the fitting of the transistor curves to the above mentioned model, and on an electrical series connected capacitor model for the device gating circuit, we explained this shift by the changes in the liquid electrical double layer formed at the graphene/solution interface. Finally, we designed a printed circuit board where the graphene chip is easily plugged in, providing a simple, robust and portable solution in view of a platform for point-of-care or other chemical and biosensing applications.

## Acknowledgements

N.C.S.Vieira acknowledges a Postdoctoral fellowship at INL from FAPESP – SP/Brazil (2014/01663-6). G. Machado Jr. acknowledges a PhD grant (no. 237630/2012-5) from CNPq – Brazil. The authors thank Clarissa Towle for proofreading the manuscript and correcting the English.

# Supplementary Information

# Graphene field-effect transistor array with integrated electrolytic gates scaled to 200 mm

N C S Vieira[1,3], J Borme[1], G Machado Jr.[1], F Cerqueira[2], P P Freitas[1], V Zucolotto[3], N M R Peres[2] and P Alpuim[1,2]

[1]INL - International Iberian Nanotechnology Laboratory, 4715-330, Braga, Portugal.
[2]CFUM - Center of Physics of the University of Minho, 4710-057, Braga, Portugal.
[3]IFSC - São Carlos Institute of Physics, University of São Paulo, 13560-970, São Carlos-SP, Brazil

## 1. Experimental

### 1.1. Graphene synthesis and characterization

Single-layer graphene (SLG) was grown by chemical vapor deposition in a load-locked quartz tube 3-zone furnace (FirstNano EasyTube® 3000) onto 99.999% purity copper (Alfa Aesar) foils (25 μm thickness and ca. 25 × 25 mm in size). A gaseous mixture of methane/hydrogen at a gas flow rate ratio of (300 sccm of H2)/(50 sccm of CH4) was used for growth. The deposition was done as follows: after transferring the copper substrate into the reactor chamber, initial heating of the catalyst takes place at 1020 ºC for 20 minutes in a H2 atmosphere, for cleaning, increasing the grain size, and surface smoothing of the copper. Flow of the growth-precursor gas, methane, follows, keeping the hydrogen flow, for 30 minutes. Growth temperature is fixed at 1020 ºC and the pressure at 0.5 Torr. Both parameters have independent closed-looped control systems. The graphene grows on both sides of the copper foil.

For graphene transfer, a temporary poly(methyl methacrylate) (PMMA) substrate was used. PMMA was spin coated onto the top side of the graphene/Cu/graphene sample and copper was further dissolved by dipping the PMMA/graphene/Cu into a 0.5 M FeCl3 solution for 2 h. PMMA/graphene was cleaned in 2% HCl solution to remove metal precipitates and further washed in deionized water five times. PMMA/graphene films were stored in ultrapure water before transferring to the pre-patterned silicon/silicon dioxide (Si/SiO2) wafer substrate. After transfer, the sample is dried with a N2 flow that also flattens the PMMA/graphene film followed by annealing for 7 hours at 180 °C to complete the drying process. The PMMA is then removed using warm acetone.

Graphene quality, i.e. the homogeneity of the obtained graphene film after transfer is first investigated by optical images. Confocal Raman spectroscopy was used to confirm the presence of SLG.

*1.2. Fabrication of the graphene electrolyte-gated field-effect transistors (EGFET)*

A 200 mm Si (100) wafer (B-doped, 8 – 30 Ω cm, LG Siltron) with 200 nm of thermal SiO2 was cleaned by ultrasonication in acetone for 5 min, rinsed sequentially in isopropanol and deionized (DI) water (≥ 18 MΩ.cm), and then dried in a nitrogen (N2) flow. The wafer was sputter-coated with chromium (Cr, 3 nm), used as adhesive layer, and gold (Au, 30 nm). Using optical lithography and ion milling, the wafer was patterned with 280 dies of ca. 10 mm in size, comprising of source and drain contacts each with a semicircular form 75 µm in diameter (channel width, W) separated by a gap (channel length, L) of 6.25, 12.5, and 25 µm, and contacts pads to connect to external measurement equipment.

An insulating layer of aluminum oxide (Al2O3, 320 nm) was patterned by lift-off on top of the contact lines, leaving uncovered the semi-circular area (corresponding to source and drain electrodes), prepared to receive the graphene. A planar ring-shaped gate of internal diameter 200 µm (figure 1c) separated ~50 µm from source and drain contacts was integrated in the transistor array. A thin layer of Al2O3 (10 nm) was deposited on top of the integrated-gate to protect it during the further microfabrication process.

The floating PMMA/graphene films were then transferred onto different areas of the pre-patterned wafer, until the desired degree of graphene coverage was obtained.

PMMA/graphene was patterned using optical lithography and oxygen plasma etching, keeping the integrated gates protected by Al2O3, which was later removed using diluted photoresist developer AZ400K 1:4 as etching agent.

After all lithographic steps, the wafer was cut into equal rectangular chips by dicing (dicing saw DISCO DAD 3350), each containing six graphene EGFETs. Each set of graphene EGFETs was washed in acetone and ethyl-acetate and dried with N2 flow, previously to the measurements.

*1.3. Graphene EGFET electrical characterization*

Graphene EGFETs were electrically characterized in phosphate buffered saline (PBS, Sigma-Aldrich P4417) solutions used as electrolytes. PBS has a total salt concentration of 161.5 mM. Measurements were performed in a computer-automated system using a Keithley 2400 source-meter and a Keithley 4687 picoammeter with an integrated voltage source. Transfer (output) transistor curves were obtained by fixing VSD and sweeping VG, and vice versa, while measuring ISD. Measurements were taken in air and the electrolyte was placed on the transistor channel by dropping of 5-20 µL from a micropipette. Some EGFETs were also electrically characterized in NaCl solutions with different ionic strengths (1.5, 15 and 150 mM) for testing. For comparison, a conventional Au wire was used as a gate electrode in selected devices. All experiments were carried out at a controlled temperature of 21-22 °C.

**2. Raman analysis, imaging and interpretation**

Raman analysis was performed in a Confocal Raman system Witec Alpha 300R using the software WITec Project Plus for data acquisition, and WITec Project FOUR+, for computing data. The Raman spectrum of graphene is characterized by the presence of two main modes, namely the G mode at $\approx$ 1580 cm$^{-1}$ (first order in plane vibrational mode) and the 2D mode at $\approx$ 2690 cm$^{-1}$, which is a second-order overtone of a different in plane vibration (of the D mode at ≈1350 cm$^{-1}$ which corresponds to an inter-valley phonon and defect scattering). The presence of the otherwise forbidden D mode is an indication of defects in the graphene.

Raman image in Figure 3 in the main text was obtained using a Nd-Yag 532 nm line at a laser power of 1.5 mW a 100× objective with a 0.9 numerical aperture and an integration time of 1s, a XY scan was performed in a scan range of 110 μm × 90 μm, with a resolution of 110 pixels per line and 90 lines. A Raman spectrum was acquired and stored for each of the 9900 pixels. After background subtraction and cosmic ray removal, a cluster analysis was done. The cluster analysis consists in the identification of similarities in the Raman spectra of the analyzed area. In order to speed up the analysis and increase the relevance of the results, we used a filter that restricts calculations to selected spectral ranges that are used to identify graphene single and multilayers (D, G and 2D peaks were chosen). The software compares all stored Raman spectra and bins them according to their similarity. Each cluster is then represented by a Raman average spectrum calculated over all the spectra belonging to that cluster (Figure 3c). The clustering is finished when creating more sub-clusters does not change the average spectrum of the sub-cluster when compared to the parent cluster average spectrum. The three average spectra form a basis which is used to build the Raman image shown in Figure 3b. The (artificially) colored image follows from the expansion of all the spectra stored in the pixels in linear components over the three basis spectra. Each pixel color results from adding the colors of the basis spectra in an amount equal to the respective coefficients in the linear expansion of the spectrum stored in that pixel.

**3. The effect of the asymmetry in contact area in the electrolytic gate capacitance**

The EGFET gate capacitance is a series combination of two electrical double-layer capacitances plus the graphene quantum capacitance, $C_q$. The two EDLs form at the Au gate contact/electrolyte interface and at the electrolyte/graphene interface, respectively. The first of these EDLs extends over an area $A_1$ equal to the gate contact area, which is the sum of the areas of the two ring-shaped lobes (see Figure S1). This area is equal to 3.32 mm$^2$. The second EDL forms over the channel area, $A_2$, which is equal to $9.4 \times 10^{-4}$ mm$^2$, in case of a code 10 transistor ($7.5 \times 10^{-4}$ and $1.9 \times 10^{-3}$ mm$^2$ for code 5 and code 20 devices, respectively). The three capacitors in series form a voltage divider. Although these capacitances are not directly measurable we can nevertheless make reasonable assumptions in order to estimate their order of magnitude. A reasonable value for $C_q$ at a value of $V_G$ arbitrarily taken from the data in Figure 2 (main text) or 5, e.g. $V_G = 0.75$ V, is 2 µF/cm$^2$, and the Debye length, $\lambda_D$, of the EDLs in PBS, is $\lambda_D = 0.78$ nm. With these assumptions we calculate the equivalent series capacitor, $C_{eq}$, of the three capacitors:

$$C_{eq} = \frac{A_1 \times C_{EDL1} \times A_2 \times C_{EDL2} \times C_{gr}}{A_1 A_2 \times C_{EDL1} C_{EDL2} + A_2 C_{EDL2} C_{gr} + A_1 C_{EDL1} C_{gr}} \tag{S1}$$

where,

$$C_{EDL1,EDL2} = \frac{\varepsilon \varepsilon_0}{\lambda_D} \text{ and } C_{gr} = C_q \times W \times L \tag{S2}$$

In equations (S2), $\varepsilon$ is the dielectric constant of water, $\varepsilon = 80$ for $C_{EDL1}$, and, taking into account the hydrophobic gap that forms at the graphene/solution interface, we assumed arbitrarily[3] $\varepsilon = 1$, for water at the graphene interface, and used this number to calculate $C_{EDL2}$. $W$ and $L$ are channel width and length, respectively. Using equation (S1) and (S2) and the capacitive voltage divider, we compare the voltage drop at the graphene quantum capacitance for devices with different channel areas. The results confirm that, for a given gate voltage ($V_G = 0.75$ V, in this example) the voltage drop is larger for code 5 channel devices ($V_q = 0.2714721$ V), followed, in decreasing order, by code 10 ($V_q = 0.2714718$ V), and code 20 EGFETs ($V_q = 0.2714712$ V). The charge concentration per unit channel area is proportional to $V_q$, therefore GFETs with shorter channel will have their minimum conductivity point shifted towards lower voltages. The differences are minimal, of the order of some µV in this example. This is not enough to explain the much larger shifts in $V_G$ observed in Figures 2 and 5 in main text.

## 4. A portable Graphene EGFET

As a demonstration of graphene EGFETs portability and ease of use in sensing platforms we designed and fabricated a printed circuit board (PCB) to support the devices, which also provided easy, durable, and precise electrical connections to the sourcing and measuring equipment. The silicon chip containing the graphene EGFET was inserted in a Samtec MB1-120 connector, which is adequate for 0.7 mm silicon substrates. With the current design, up to three devices can be measured on each chip. A 3-pole switch selects which of the transistor circuits is addressed. Five banana connectors are provided at the PCB side opposite to the chip connection, to plug in the measuring equipment. Connections for source, drain, gate, ground, and an auxiliary port are provided.

**Supplementary Figures**

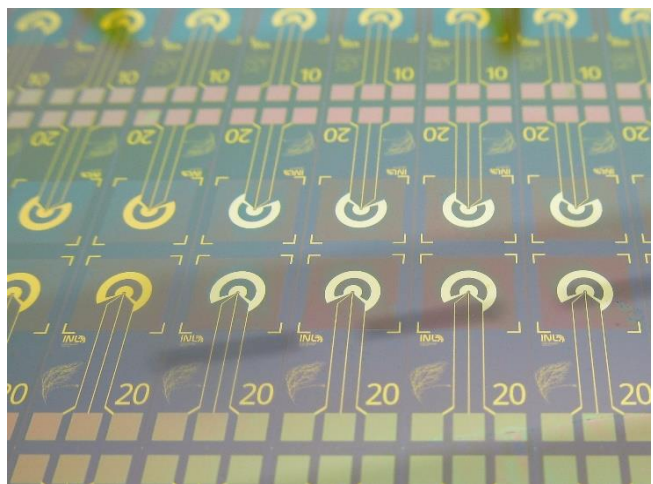

**Figure S1.** Partial view of the graphene transistor array, in an area where devices with channel length 25 µm (labelled 20), and 12.5 µm (labelled 10) are visible.

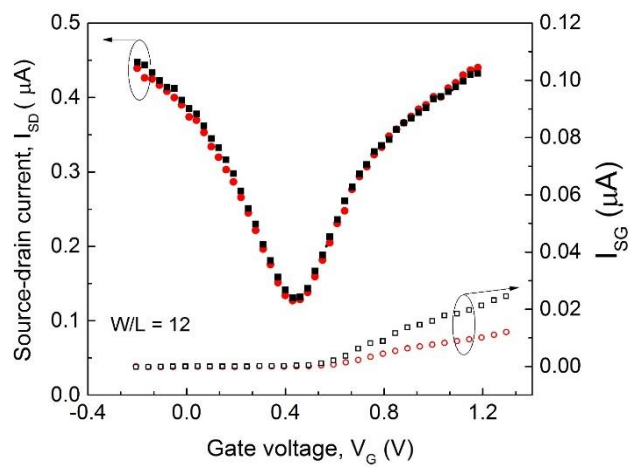

**Figure S2.** Transfer curves of a graphene EGFET gated via the integrated-gate (solid red dots) and via a conventional Au wire (solid black squares). Gate-source leakage current measured using the integrated-gate (open red dots) and the wire gate (empty black squares). Measurements performed in PBS at 25 °C.

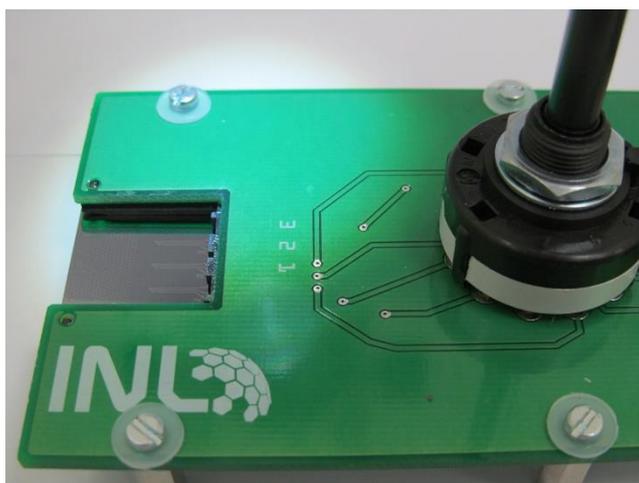

**Figure S3.** Printed circuit board designed for (a) easy plug-in of the graphene sensor, at one end, and (b) provide easy and rugged connection to electrical measuring equipment, at the other end. A 3-pole switch selects which of three different sensor circuits is addressed. The silicon chip is $25 \times 22$ mm$^2$ in size.